\documentclass[conference]{IEEEtran}
\IEEEoverridecommandlockouts

\usepackage{cite}
\usepackage{amsmath,amssymb,amsfonts}
\usepackage{algorithmic}
\usepackage{graphicx}
\usepackage{textcomp}
\usepackage{xcolor}
\usepackage{booktabs}
\usepackage{url}
\usepackage{flushend}
\usepackage{tikz}

\begin{document}

\title{A Reference Architecture for Agentic Hybrid Retrieval in Dataset Search}

\author{%
  \IEEEauthorblockN{Riccardo Terrenzi}
  \IEEEauthorblockA{\textit{Centre for Industrial Software} \\
    \textit{University of Southern Denmark}\\
    Alsion 2, S{\o}nderborg, 6400, Denmark\\
  rite@mmmi.sdu.dk}
  \and
  \IEEEauthorblockN{Phongsakon Mark Konrad}
  \IEEEauthorblockA{\textit{Centre for Industrial Software} \\
    \textit{University of Southern Denmark}\\
    Alsion 2, S{\o}nderborg, 6400, Denmark\\
  phkon23@student.sdu.dk}
  \and
  \IEEEauthorblockN{Tim Lukas Adam}
  \IEEEauthorblockA{\textit{Centre for Industrial Software} \\
    \textit{University of Southern Denmark}\\
    Alsion 2, S{\o}nderborg, 6400, Denmark\\
  tiada23@student.sdu.dk}
  \and
  \IEEEauthorblockN{Serkan Ayvaz}
  \IEEEauthorblockA{\textit{Centre for Industrial Software} \\
    \textit{University of Southern Denmark}\\
    Alsion 2, S{\o}nderborg, 6400, Denmark\\
  seay@mmmi.sdu.dk}
}

\maketitle

\begin{abstract}
  Ad hoc dataset search requires matching underspecified natural-language queries against sparse, heterogeneous metadata records, a task where typical lexical or dense retrieval alone falls short. We reposition dataset search as a software-architecture problem and propose a bounded, auditable reference architecture for agentic hybrid retrieval that combines BM25 lexical search with dense-embedding retrieval via reciprocal rank fusion (RRF), orchestrated by a large language model (LLM) agent that repeatedly plans queries, evaluates the sufficiency of results, and reranks candidates. To reduce the vocabulary mismatch between user intent and provider-authored metadata, we introduce an offline metadata augmentation step in which an LLM generates pseudo-queries for each dataset record, augmenting both retrieval indexes before query time. Two architectural styles are examined: a single ReAct agent and a multi-agent horizontal architecture with Feedback Control. Their quality-attribute tradeoffs are analyzed with respect to modifiability, observability, performance, and governance. An evaluation framework comprising seven system variants is defined to isolate the contribution of each architectural decision. The architecture is presented as an extensible reference design for the software architecture community, incorporating explicit governance tactics to bound and audit nondeterministic LLM components.
\end{abstract}


\section{Introduction and Motivation}

Although open data portals contain millions of dataset records, identifying a dataset that precisely addresses a specific research question remains challenging. For example, a query for ``COVID hospitalization rates by county'' may yield no results for a dataset titled ``US Health Statistics 2020, County-Level Hospital Admissions.'' This persistent barrier stems from a mismatch between user intent and provider-authored metadata~\cite{noy2019}. In addition to vocabulary mismatch, dataset metadata frequently lacks detail: titles are often terse, descriptions are incomplete, and tags are inconsistent across different portals~\cite{castelo2021}.

These issues are not merely usability problems: they systematically degrade retrieval quality in settings where the correct dataset is present but not lexically discoverable under the user’s query. Traditional dataset search pipelines typically assume a single-pass formulation (query $\rightarrow$ retrieve $\rightarrow$ rank) \cite{chapman2019datasetsurvey}, placing the burden of query refinement on the user and relying on portal-specific indexing choices. However, dataset search is inherently iterative and multi-faceted: users often need to reconcile synonyms, units, temporal and geographic scope, population definitions, and licensing constraints before a dataset can be judged relevant \cite{paton2024survey}. When metadata is sparse, even strong neural retrieval models may fail because the evidence required to establish relevance is simply absent from the indexed record~\cite{paton2024survey, gan2025keywords_not_always_key, chapman2019datasetsurvey}.

Current dataset search engines attempt to solve these challenges at the system level. Google Dataset Search~\cite{noy2019} aggregates schema.org\footnote{\url{https://schema.org}} markup across the web, while Auctus~\cite{castelo2021} combines profiling with keyword search. However, neither system incorporates iterative retrieval procedures or semantic matching techniques.

The rise of LLM-based agents gives the opportunity to shift dataset search from a single-pass ranking problem to an iterative, tool-orchestrated workflow in which the system plans, retrieves, evaluates evidence, and revises queries over multiple steps \cite{sun2023}\cite{lu2024responsible_refarch}. This shift is primarily architectural: it introduces a control loop, tool boundaries, and new governance requirements (e.g., bounded cost/latency, observability, reproducibility, and auditability).

To this end, we introduce an agent-based reference architecture for dataset search. The proposed architecture elevates retrieval from a single-stage function to a bounded, iterative workflow structured around a Plan--Retrieve--Evaluate loop. An LLM-based agent decomposes the user request, orchestrates hybrid retrieval (sparse and dense) through reciprocal rank fusion (RRF), evaluates intermediate results, and selectively rewrites queries under an explicit iteration budget. This control loop makes refinement systematic rather than user-dependent, while preserving observability and bounded execution.

Beyond the single-agent design, we further propose (i) an offline metadata augmentation strategy based on LLM-generated pseudo-queries to mitigate first-pass retrieval failures caused by incomplete metadata, and (ii) a multi-agent decomposition in which specialized agents interact through typed contracts under feedback control, enabling higher performances, clearer responsibility boundaries and improved governability.

This paper offers three primary contributions:
\begin{enumerate}
  \item A single-agent reference architecture for dataset search that fuses sparse and dense retrieval via reciprocal rank fusion (RRF). The architecture is governed by an LLM agent with an iterative Plan--Retrieve--Evaluate loop (Section~III).

  \item An offline metadata augmentation strategy using LLM-generated pseudo-queries to reduce first-pass retrieval failures (Section~IV).

  \item A multi-agent decomposition, implemented as a specialized agent pipeline with a feedback loop, together with typed inter-agent contracts. We also provide a quality-attribute tradeoff analysis and a governance checklist for bounding, auditing, and reproducing LLM-driven retrieval (Sections~V and~VI).
\end{enumerate}

Our hypotheses are: (i) the single-agent architecture outperforms retrieval-only baselines in terms of ranking metrics (i.e. NDCG), demonstrating that iterative architectural control yields measurable gains over static pipelines; (ii) the offline pseudo-query augmentation reduces first-pass retrieval failures and decreases the number of iterations required to reach high-quality results; (iii) the multi-agent decomposition exceeds the retrieval performances of the single-agent architecture and improves modularity, observability, failure isolation, and governance.

The objective of this study is to reposition dataset search as an architectural design problem rather than solely an information retrieval problem, and provide a reproducible reference architecture for building bounded, auditable, and evolvable LLM-driven retrieval systems.

\section{Related Work}

\subsubsection{Dataset Search and Metadata-Centric Retrieval}

Recent dataset discovery work increasingly treats metadata as the main retrieval substrate, driven by scalability and access constraints. Paton et al.\ survey the dataset discovery design space and show that, even at portal scale, search quality is often limited more by sparse, heterogeneous metadata than by corpus size~\cite{paton2024survey}. Human-in-the-loop evidence likewise shows persistent failures from vocabulary mismatch, ambiguous intent, and missing metadata context, motivating iterative, interaction-aware retrieval over single-shot keyword search~\cite{hulsebos2024hilda}. Recent systems address this with richer metadata-centric representations: Metam models goal-oriented discovery by operationalizing user goals through intermediate reasoning over dataset descriptors and candidate transformations~\cite{galhotra2023metam}, while BLEND frames data discovery as an end-to-end pipeline unifying tasks such as entity- and attribute-centric lookup, again stressing robust metadata acquisition and integration as prerequisites for effective search~\cite{esmailoghli2025blend}.

\subsubsection{Hybrid Retrieval and Retrieval-Augmented Generation}

Recent RAG surveys describe a shift from ``na\"ive'' pipelines to modular designs that separately optimize retrieval, evidence selection, and generation~\cite{gao2023rag_survey}. On the retrieval side, sparse (lexical) and dense (semantic) signals are widely viewed as complementary, motivating fusion strategies such as RFF. Yang et al.\ propose sparse-guided partial dense retrieval, which evaluates only a subset of embedding clusters to preserve fusion quality while lowering dense retrieval cost~\cite{yang2024cluster_partial_dense}. A second line of work improves retrieval through query adaptation and diversification: Rewrite--Retrieve--Read casts RAG as query rewriting and shows that reformulating the input can substantially improve downstream results~\cite{ma2023query_rewriting_emnlp}, while RAG-Fusion generates multiple queries and applies reciprocal-rank-style fusion to improve recall and coverage under paraphrase and ambiguity~\cite{rackauckas2024rag_fusion}. A third line adds control policies for when and how to retrieve: Self-RAG learns retrieval-on-demand and uses self-critique signals to regulate evidence use during generation~\cite{asai2024selfrag}.

\subsubsection{Agentic Architectures and LLM-Orchestrated Systems}

Recent work treats LLM agents as compound software systems whose reliability depends on orchestration, interfaces, and governance. The Agent Design Pattern Catalogue synthesizes recurring patterns and tradeoffs for foundation-model-based agents, enabling more principled decomposition and observability~\cite{liu2025agentpatterns}. Moving from guidance to evaluation, AgentArcEval adapts scenario-based architecture assessment to FM-based agents and targets quality attributes that model-only metrics often miss, such as autonomy and continuous evolution~\cite{lu2025agentarceval}. Multi-agent orchestration is studied for both capability and engineering control: Mixture-of-Agents formalizes layered collaboration across multiple LLMs~\cite{wang2025moa}, while Gradientsys introduces a scheduler that routes tasks among specialized agents via typed interfaces and explicit replanning under failure~\cite{song2025gradientsys}. Governance is also becoming runtime infrastructure; MI9 emphasizes continuous monitoring and conformance checks for agentic behavior beyond pre-deployment controls~\cite{wang2025mi9}. Finally, interactive benchmarks make architectural claims testable: X-WebAgentBench evaluates multilingual web-agent planning and interaction~\cite{wang-etal-2025-x}, and LoCoBench-Agent extends long-context software engineering evaluation to multi-turn, tool-using workflows with efficiency and recovery metrics~\cite{qiu2025locobenchagent}.
\section{Approach: Agentic Hybrid Retrieval}

We address the problem of ad-hoc dataset search for tabular data. Given a collection of tables $\mathcal{D}=\{D_1,\dots,D_M\}$ and a natural language query $q$, the goal is to return a ranked list of relevant datasets.

\begin{figure}[t]
\centering
\includegraphics[width=0.95\linewidth]{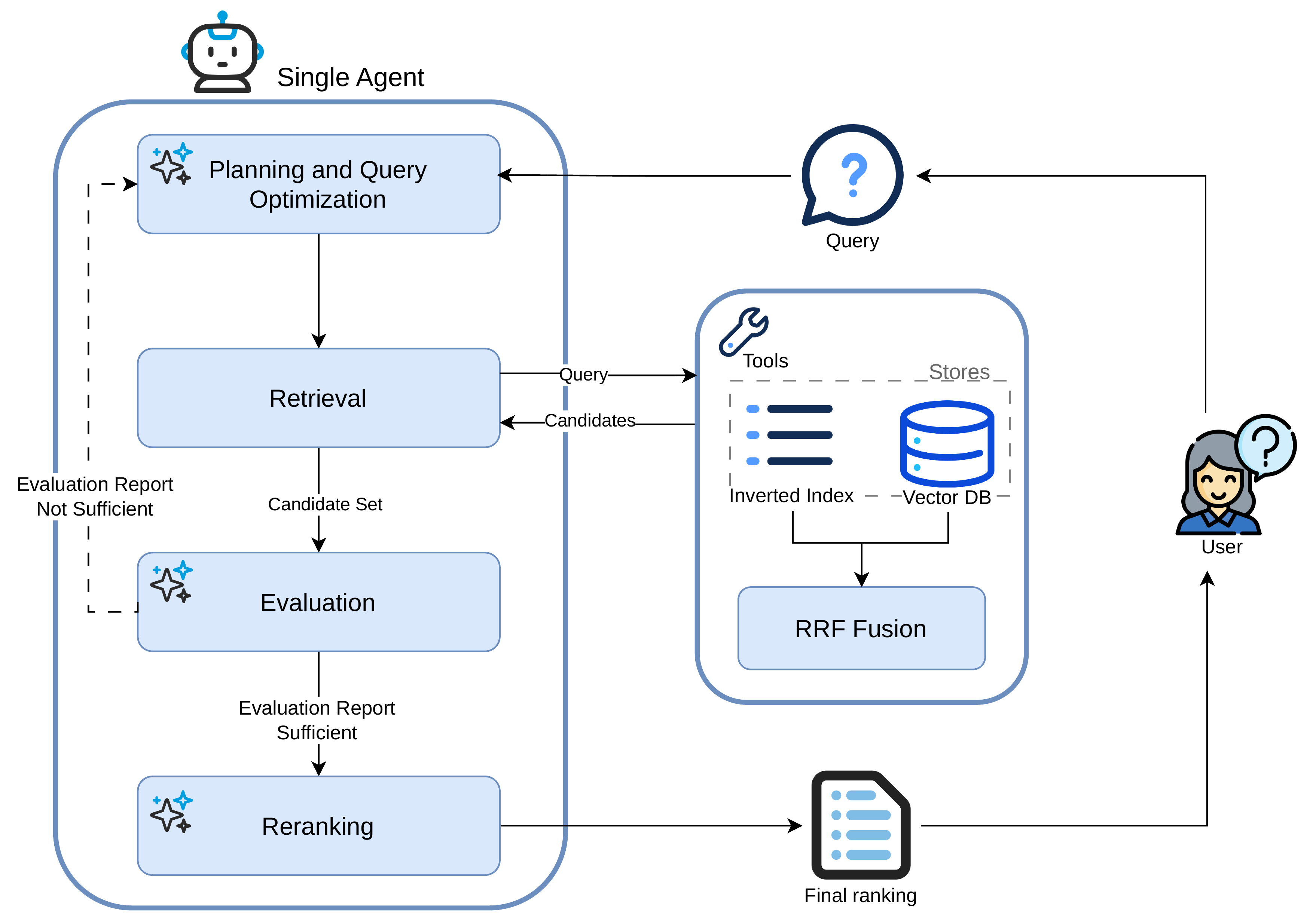}
\caption{Single Agent with Plan--Retrieve--Evaluate loop. Dashed arrows denote feedback loops. }
\label{fig:single_agent}
\end{figure}

We frame the proposed architecture as a single-agent, iterative hybrid RAG architecture with a Plan-Retrieve-Evaluate control loop: a single LLM agent controls a hybrid retriever over dataset metadata and performs listwise relevance reasoning over the retrieved candidates. The proposed architecture is shown in Fig.~\ref{fig:single_agent}.

Moreover, we adopt dataset metadata as the primary searchable representation, since most dataset search in open data portals operates over publisher-provided metadata and typically expose keyword search and faceted metadata filtering rather than direct content-based search~\cite{chapman2019datasetsurvey,noy2019, paton2024survey}. This setting is particularly challenging because metadata is often sparse, inconsistently defined across portals, and authored without a shared vocabulary, amplifying ambiguity and vocabulary mismatch at query time~\cite{chapman2019datasetsurvey, paton2024survey}.

At a high level, the architecture consists of a single LLM agent, two complementary metadata stores and one retrieval tool:
\begin{itemize}
    \item Single LLM agent: an agent responsible for query planning, retrieval, candidates evaluation and candidates listwise reranking.
    \item Lexical metadata index: an inverted index built over serialized dataset metadata, supporting classical lexical matching (BM25).
    \item Vector database: a vector store containing dense embeddings of the same serialized metadata, enabling semantic retrieval.
    \item Retriever tool: executes lexical and dense retrieval and returns a fused candidate set with provenance.
\end{itemize}

Concretely, instead of delegating query planning, retrieval evaluation and reranking to separate autonomous components, the single agent performs these steps autonomously within its own decision loop, conditioning on the retrieved observations. This design choice is motivated by recent work showing that agentic RAG-style controllers can improve robustness by iterating between retrieval and self-critique, rather than relying on a single retrieve-then-rank step~\cite{liu2024raisf,yu2024autorag,craftingPath2024}.
Moreover, in the dataset domain, benchmark collections such as the NTCIR-derived ad hoc dataset retrieval test collection~\cite{kato2021adhoc} and ACORDAR~2.0~\cite{acordar2} emphasize that successful dataset retrieval often requires semantic interpretation beyond simple term matching. This evidence supports the decision to use a hybrid, iterative retrieval phase.

\subsection{Data Stores population}

The metadata catalogue content is used to populate both a BM25 inverted index and a VectorDB.
It is usually composed of entries like this one:
\begin{verbatim}
{
  {
    "download": "...",
    "size": "...",
    "author": "...",
    "created": "...",
    "dataset_id": "...",
    "description": "...",
    "title": "...",
    "version": "...",
    "tags": "..."
  },
  ...
}
\end{verbatim}

In order to populate the inverted index and the vector database, each dataset record is serialized as a flat text string with the following fields. We select these fields because they carry most of the lexical and semantic evidence available for content-based dataset search: titles and descriptions capture topical meaning and paraphrases, tags provide high-precision keywords, and provider/author fields often include distinctive named entities that help disambiguate similar datasets.
\begin{verbatim}
Title is {...}, Description is {...},
Tags are {...}, Author is {...}
\end{verbatim}

\subsection{Single agent workflow}

The workflow begins when the system receives a natural-language user query. The agent will follow a Plan--Retrieve--Evaluate cycle, inspired by ReAct-style reasoning~\cite{yao2023}. The agent first analyzes the query and decides how to optimize it for retrieval, e.g., by decomposing it into sub-questions, diversifying it into multiple complementary formulations, and determining whether a rewrite is necessary to reduce vocabulary mismatch. Based on this analysis, the agent builds a query plan specifying how many candidates to fetch per channel, and which rewritten or expanded queries to issue.

Given a query plan, the agent uses the retrieval tool, which searches both indexes and merges the results using reciprocal rank fusion (RRF)~\cite{cormack2009}:
\[
  \mathrm{RRF}(d) = \sum_{r} \frac{w_r}{k + \mathrm{rank}_r(d)} \, ,
\]
with configurable weights.
where:
\begin{itemize}
  \item $d$ is a candidate dataset record (document);
  \item $r$ indexes a retrieval channel (e.g., BM25 or dense);
  \item $\mathrm{rank}_r(d)$ is the rank position of $d$ returned by channel $r$ (lower is better);
  \item $w_r$ is the weight assigned to channel $r$;
  \item $k$ is a smoothing constant controlling the contribution of lower-ranked results.
\end{itemize}
Depending on the metadata context, it is possible to assign greater or lesser weight to one retrieval channel: the BM25 channel emphasizes lexical overlap and tends to work best when metadata contains distinctive identifiers, codes, or standardized terms, whereas the dense channel emphasizes semantic similarity and is more robust when metadata is short, noisy, or expressed via synonyms and paraphrases~\cite{bruch2023analysis}.

The tool returns the merged candidate list to the agent, which evaluates the candidates based on the similarity score, the variety and diversity of the results, and the genericity of the answer. In this setup, similarity score is the only objective metric. Criteria such as diversity, variety, and genericity are heuristically assessed by the autonomous LLM, without additional evaluation tools.
At each iteration, the agent may:
\begin{itemize}
\item rewrite the query to address vocabulary gaps identified in the current candidates;
\item issue additional retrieval calls with refined terms;
\item declare the candidate set sufficient.
\end{itemize}
Upon sufficiency, the agent performs a listwise rerank~\cite{sun2023} over the top-$N$ candidates and emits the final ranking. This looping architecture distinguishes it from single-pass RAG pipelines, as the controller adapts its strategy based on intermediate results rather than relying on a single query formulation.

\subsection{Governance and bounded execution}
Since the proposed system introduces an explicit control loop, we treat governance as a first-class architectural concern. Each run is bounded by an iteration budget $T_{\max}$, a maximum number of tool calls, and per-channel top-$K$ limits to cap cost and latency, with a deterministic fallback that returns the best candidate set observed so far when budgets are exhausted. For observability and auditability, every iteration emits a structured provenance trace capturing: the agent’s query rewrites and plan, retrieval parameters (BM25/dense settings, $w_r$, $k$), per-channel ranked lists with scores, the fused RRF list, and the evaluation rationale used to stop or continue. All artifacts are versioned (model, prompts, retriever configuration) to enable replay-based reproducibility and to localize failures to a specific step (planning, retrieval, fusion, or reranking), supporting debugging and controlled evolution of the architecture.

\section{Metadata Augmentation}

\begin{figure}[t]
\centering
\includegraphics[width=0.95\linewidth]{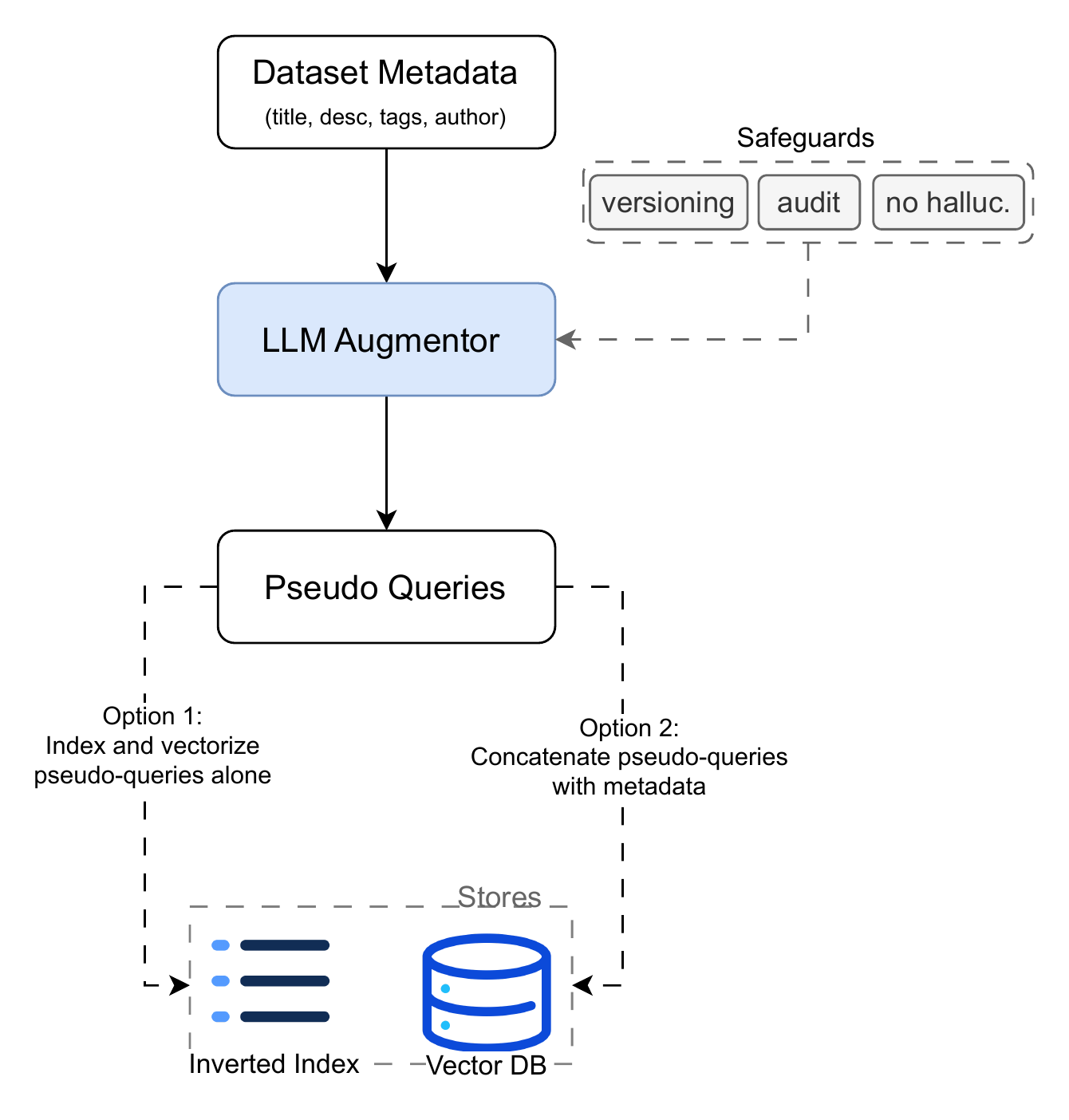}
\caption{Offline metadata augmentation pipeline. The LLM Augmentor generates pseudo-queries from dataset metadata, subject to versioning and audit safeguards.}
\label{fig:augmentation}
\end{figure}

A core limitation of any retrieval system is the quality of its indexed content. Dataset metadata is authored by data providers with no knowledge of how future users will search, creating a systematic gap between the index and query vocabularies~\cite {gan2025keywords_not_always_key, zhang2025autoddg}. This limitation is addressed by introducing an offline augmentation step, as shown in Fig.~\ref{fig:augmentation}. For each dataset record, an LLM generates a set of 5--10 natural-language pseudo-queries that a user may plausibly issue when seeking this dataset.

The augmentation step is proposed as an offline step to improve the perfomances of the single agent architecture: having a more convenient retrieval leads to less iterations needed. This approach draws on the idea of improving table retrieval using question generation \cite{liang2025improving} and hypothetical document embeddings~\cite{gao2022hyde}, adapted to leverage LLM generative capacity rather than corpus-based feedback. The rationale follows a shift-left strategy: improving the index offline reduces the number of iterative query rewrites required at query time, thereby improving both latency and result quality during the initial retrieval pass. Pseudo-elements are generated once per dataset record and amortized across all future queries, making the additional LLM cost a fixed offline investment rather than a recurring per-query expense.

The generated pseudo queries can be employed in two different ways that will be evaluated as ablation studies:

\begin{itemize}
    \item The pseudo-queries are indexed in a dedicated BM25 index and vector store, and are directly used for retrieval instead of the original indexed metadata. Each pseudo-query includes metadata describing the dataset from which it is derived. 
    \item The pseudo-queries are concatenated alongside the original metadata and re-indexed into the BM25 and vector stores.
\end{itemize}

\subsection{Safeguards}

To maintain trustworthiness, the augmentation process is constrained: (i)~generated content is derived only from existing metadata (the LLM may not hallucinate external facts); (ii)~when metadata is insufficient, the model outputs ``unknown'' for that facet; (iii)~each augmented record stores the model identifier and prompt version for reproducibility; and (iv)~a configurable sampling audit manually verifies a random subset of generated pseudo-queries against source metadata.

\section{Multi-Agent Decomposition}

\begin{figure}[t]
\centering
\includegraphics[width=0.95\linewidth]{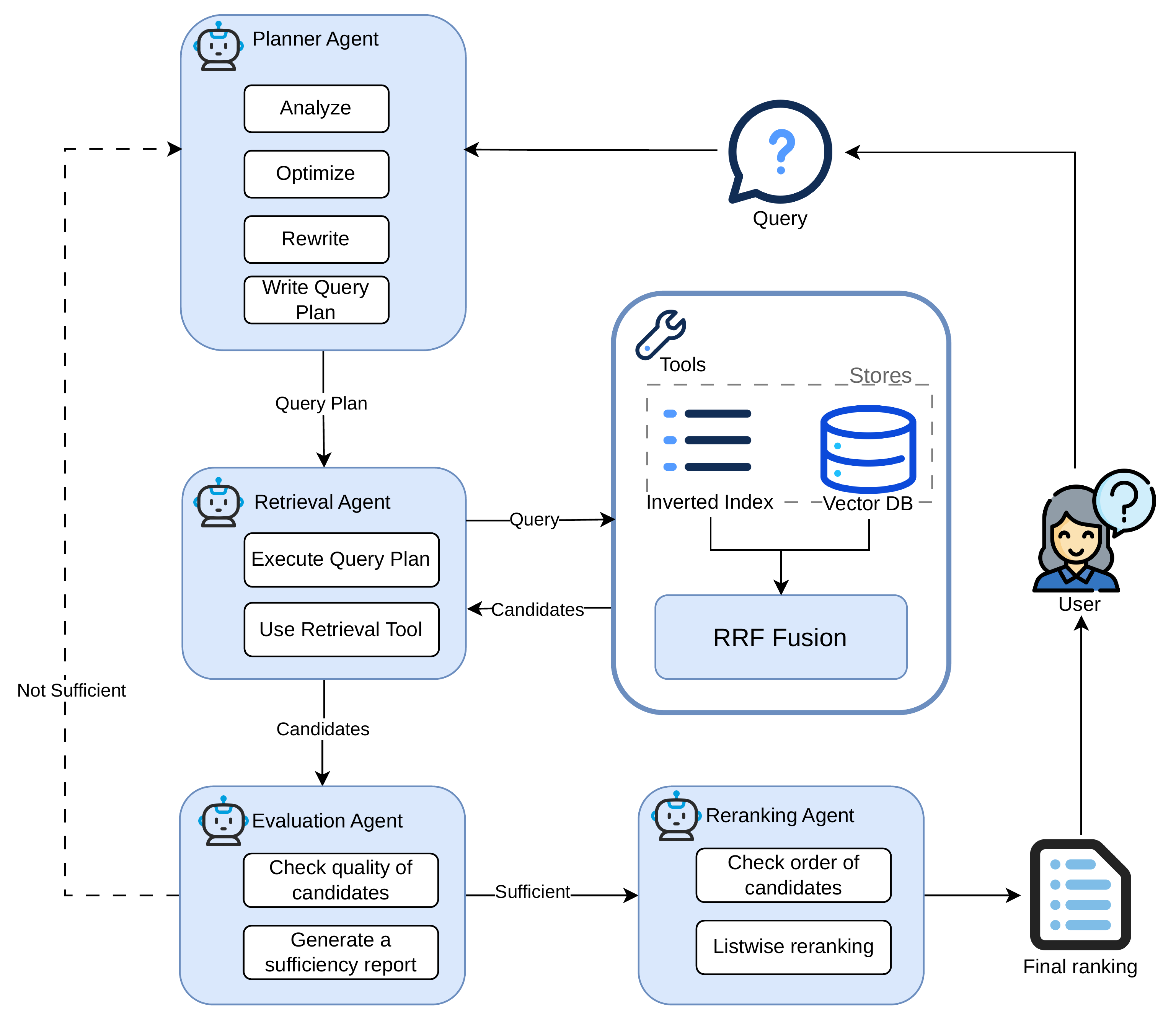}
\caption{Multi Specialized-Agent Pipeline. Edge labels are typed inter-agent contracts.}
\label{fig:multi_agent}
\end{figure}

The single-agent architecture centralizes all reasoning in a single LLM agent. As retrieval tasks become more complex, such as multi-faceted queries requiring geographic and temporal filtering, maintaining and testing a monolithic prompt becomes increasingly challenging. We therefore propose a multi-agent decomposition, shown in Fig.~\ref{fig:multi_agent}, that factors the single agent into four specialized agents.

Recent results also indicate that strong single-agent systems can remain competitive when endowed with capability abstractions (skills/tools) and sufficiently large context windows to preserve coherent state across multi-step execution~\cite{li2026sas_vs_mas}. In our setting, we hypothesize that as dataset search tasks become more complex (e.g., requiring multi-faceted decomposition, iterative filtering, and tool-intensive reasoning), switching to a multi-agent architecture can yield superior performance by enabling structured collaboration and iterative refinement across specialized roles~\cite{wang2025moa}.

\subsection{Multi-agent workflow}

In this case, instead of a single agent switching roles, each role is implemented as a dedicated agent with a well-defined responsibility.
The Planner Agent receives the user query, analyzes it, and decides how to optimize it (e.g., by decomposing it into subqueries or adding key terms). The Planner then outputs a query plan which specifies what to search by enumerating the concrete query strings to issue (including expansions and decompositions) and any high-level constraints, but it does not encode low-level retrieval hyperparameters. The Retriever Agent consumes the query plan and autonomously determines how retrieval is executed, including the number of candidates retrieved per channel (lexical vs.\ dense) and channel weighting. It then executes the resulting retrieval procedure against BM25 and vector stores, producing a candidate set: a list of dataset metadata comprising title, description, tags, and author. The Evaluator Agent scores the candidate set based on the similarity score, the variety and diversity of results, and the genericity of the answer. Then it generates an evaluation report that states whether the candidate set is sufficient and explains the reasons. If not sufficient, the Evaluator triggers the Planner to evaluate again the query in relation to the evaluation report, resulting in a feedback loop. If sufficient, the Evaluator triggers the Reranking agent, which receives the candidate set, checks its ordering and performs a listwise reranking of the set. The final ranking is a ranked list of datasets, each paired with its relative details, and is submitted to the user.

Each inter-agent artifact (query plan , candidate set, evaluation report, final ranking, ) is defined by a strict JSON schema. This makes the pipeline composable, independently testable, and observable, as each agent can be swapped, versioned, or mocked without affecting others.

\subsection{Quality-attribute tradeoffs}

The multi-agent style enhances modifiability by permitting independent replacement of agents, strengthens governance through per-agent logging and typed contracts for auditability, and improves failure isolation, as a faulty reranker does not compromise the retrieval plan. However, this approach incurs coordination costs, increases the number of LLM calls, and introduces new failure modes, including chain failures and schema drift across agent versions. A summary of the architectural tradeoffs is presented in Table~\ref{tab:tradeoffs}.

\begin{table}[t]
\caption{Quality-attribute trade-offs between single-agent and multi-agent architectures.}
\label{tab:tradeoffs}
\centering
\small
\renewcommand{\arraystretch}{1.15}
\setlength{\tabcolsep}{5pt}
\begin{tabular}{p{0.44\columnwidth}cc}
\hline
\textbf{Quality attribute} & \textbf{Single-agent} & \textbf{Multi-agent} \\
\hline
Modifiability, Testability & M & H \\
Governance, Observability  & M & H \\
Performance--Cost ratio    & H & M \\
Failure isolation          & L & H \\
Operational complexity     & L & H \\
\hline
\end{tabular}

\vspace{0.35em}
\footnotesize
\textit{Note:} H = High; M = Moderate; L = Low.
\end{table}

Risks are mitigated via a layered set of governance and robustness tactics. First, strict JSON schema validation can be enforced at each agent boundary, ensuring that malformed outputs and contract drift are detected at the point of production rather than propagating downstream. Second, the multi-agent pipeline benefits from an explicit separation of concerns between planning and execution, which limits cross-agent coupling and clarifies responsibilities. To preserve repeatability and facilitate causal attribution in ablation and regression analyses, agent autonomy can be bounded by enforcing fixed parameters and logging the runtime-selected parameters. Third, caching can be applied: once a Planner emits a query plan, candidate set, evaluation report and final ranking are cached and keyed by a cryptographic hash of the plan and the index version, thereby reducing redundant LLM calls under repeated executions. Finally, a global iteration budget, both a maximum number of refinement cycles and a wall-clock timeout, bounds cost and precludes non-terminating agent loops. 

\section{Evaluation Design}

An ablation study is planned out across seven system variants to isolate the contribution of each architectural component, as shown in Table~\ref{tab:variants}. Variants V1 and V2 are single-signal baselines, where V1 is based on sparse BM25 and V2 on dense retrieval. V3 combines both signals via Reciprocal Rank Fusion (RRF) to quantify the benefit of hybrid retrieval without any LLM-driven reasoning. V4 introduces the single-agent setup that performs planning, retrieval, evaluation, and reranking. V5 extends V4 with pseudo-query generation to diversify the search space and reduce the number of refinement iterations needed. V6 mirrors V4 but decomposes the single agent into a multi-agent pipeline (planner, retriever, evaluator, reranker) to improve accuracy, observability and failure isolation. Finally, V7 mirrors V5 in the multi-agent setting by adding pseudo-queries on top of the multi-agent pipeline.

\begin{table}[t]
\caption{Evaluation variants: component matrix.}
\label{tab:variants}
\centering
\small
\renewcommand{\arraystretch}{1.15}
\setlength{\tabcolsep}{7pt}
\begin{tabular}{lccccccc}
\hline
\textbf{Component} & \textbf{V1} & \textbf{V2} & \textbf{V3} & \textbf{V4} & \textbf{V5} & \textbf{V6} & \textbf{V7} \\
\hline
BM25           & P & --- & P & P & P & P & P \\
Dense          & --- & P & P & P & P & P & P \\
RRF            & --- & --- & P & P & P & P & P \\
Planner        & --- & --- & --- & I & I & P & P \\
Retriever      & --- & --- & --- & I & I & P & P \\
Evaluator      & --- & --- & --- & I & I & P & P \\
Reranker       & --- & --- & --- & I & I & P & P \\
Pseudo-queries & --- & --- & --- & --- & P & --- & P \\
\hline
\end{tabular}

\vspace{0.35em}
\footnotesize
\textit{Note:} P = present; I = implemented inside the single-agent controller.
\end{table}

\subsection{Retrieval and architectural metrics}

The selected metrics are the classic ones for information retrieval systems: Normalized Discounted Cumulative Gain (nDCG)~\cite{jarvelin2002}, Mean Average Precision (MAP), Recall, and Mean Reciprocal Rank (MRR), evaluated at $k \in \{5, 10, 20\}$.

Beyond retrieval quality, we measure iterations per query, total tool calls, end-to-end latency, token consumption per query, result steadiness across repeated runs (to quantify LLM nondeterminism), and schema validation failure rate (multi-agent only). Stability is operationalized as the Jaccard similarity of top-$k$ result sets across five independent runs with identical inputs, providing a direct measure of the nondeterminism introduced by LLM components.

Architectural qualities are operationalized as follows: modularity via component coupling/cycles and change-impact size (\#modules/contracts touched) under controlled refactor tasks; observability via trace/log completeness and time-to-root-cause in injected incident scenarios; failure isolation via blast radius, partial-result availability, and $\Delta$nDCG under fault injection; and governance via provenance completeness (audit replayability) and schema/contract compliance rate (multi-agent only).

\subsection{Large Language Models and Datasets}

To carry out the evaluation, prototypes will be developed for both architectures (single- and multi-agent). The prototypes will be based on state-of-the-art (SOTA) models, both open-source and closed-source. For closed-source models, we considered OpenAI GPT 5.2\footnote{\url{https://openai.com/}}, Anthropic Opus 4.6\footnote{\url{https://www.anthropic.com/}}, and Google Gemini 3 Deep Think\footnote{\url{https://www.google.com/}}. For open-source models, we considered Kimi K2.5\footnote{\url{https://www.kimi.com/}} and Alibaba Qwen3.5\footnote{\url{https://qwen.ai}}.
In addition, the multi-agent architecture will also be implemented with more economical mid-range models to compare performance and costs with the single-agent system equipped with a SOTA model. Models for this implementation include Anthropic Sonnet 4.6, OpenAI gpt5-mini, and Alibaba Qwen3-coder-next.

The primary benchmark is the ACORDAR 2.0 test collection~\cite{acordar2}, a large ad hoc dataset-retrieval benchmark with question-style queries and graded relevance judgments over real-world open data portals. We evaluate both its natural-language and keyword queries, and report results on the full corpus as well as a stratified 500-record subset spanning multiple domains. Additional evaluations use NTCIR-15 Data Search~\cite{kato2021adhoc} and the TARGET table-retrieval benchmark~\cite{target2024} as a stress test of generalization beyond classic ad hoc dataset search.

\section{Conclusion}

This work presents a reference architecture for agentic hybrid retrieval in ad hoc dataset search, combining BM25 and dense retrieval under an LLM-orchestrated controller loop. Offline metadata augmentation using pseudo-queries addresses the vocabulary mismatch between user intent and provider-authored metadata without incurring per-query LLM costs. A multi-agent decomposition using typed contracts provides an alternative architectural style when governance, modifiability, and independent testability are emphasized over latency and operational simplicity. A seven-variant ablation framework isolates the contribution of each architectural layer, providing a systematic basis for empirical comparison of single-agent and multi-agent design trade-offs.

\bibliography{references}

\end{document}